%% file: main.tex
\documentclass{llncs}

\usepackage{booktabs} 

\usepackage{amsfonts, amsmath}
\usepackage{graphicx,color}
\usepackage{listings}
\usepackage{pifont}
\usepackage{wrapfig}
\usepackage{tikz}
\usepackage{url}
\usepackage{ wasysym }
\usetikzlibrary{shapes,arrows}
\usepackage[linesnumbered,ruled]{algorithm2e}
\usepackage{multicol}

\usepackage[colorinlistoftodos,prependcaption,textsize=tiny]{todonotes}

\input ztmpops
\input macros

\title{Adaptive Testing for Specification Coverage}
\author{Ezio Bartocci\inst{1} \and Roderick Bloem\inst{2} \and Benedikt Maderbacher\inst{2} \and \\ Niveditha Manjunath\inst{1,3} \and Dejan Ni\v{c}kovi\'{c}\inst{2}}
\institute{Vienna University of Technology \and Graz University of Technology \and AIT Austrian Institute of Technology}

\begin{document}

\maketitle


\begin{abstract}
	Ensuring correctness of cyber-physical systems (CPS) is a challenging task that is in practice often addressed with simulation-based testing. Formal specification languages, such as Signal Temporal Logic (STL), are used to mathematically express CPS requirements and thus render the simulation activity more principled. We propose a novel method for adaptive generation of tests with specification coverage for STL. To achieve this goal, we devise cooperative reachability games that we combine with numerical optimization to create tests that explore the system in a way that exercise various parts of the specification. To the best of our knowledge our approach is the first adaptive testing approach that can be applied directly to MATLAB\texttrademark\; Simulink/Stateflow models. We implemented our approach in a prototype tool and evaluated it on several illustrating examples and a case study from the avionics domain, demonstrating the effectiveness of adaptive testing to (1) incrementally build a test case that reaches a test objective, (2) generate a test suite that increases the specification coverage, and (3) infer what part of the specification is actually implemented. 
\end{abstract}


\input intro

\input related

\input background

\input games

\input casestudy

\input conc


\bibliographystyle{plain}
\bibliography{biblio} 

\end{document}

%% file: ztmpops.tex
\font\zmsy=zmsy10

\newcommand\dia{\mathop{\hbox{\zmsy\char'061}}}
\newcommand\boc{\mathop{\hbox to 11truept{\hfill\zmsy\char'060\hfill}}}

\newcommand\cirM{\mathop{\hbox{\zmsy\char'023}}}
\newcommand\diaM{\mathop{\hbox{\zmsy\char'121}}}
\newcommand\boxM{\mathop{\hbox to 11truept{\hfill\zmsy\char'140\hfill}}}

\newcommand\until{\hskip 1.5pt\hbox{$\mathcal{U}$}\hskip 0.5pt}
\newcommand\since{\hskip 1.5pt\hbox{$\mathcal{S}$}\hskip 0.5pt}

\newcommand\Previous{\cirM}

\newcommand\Glob{\boc}
\newcommand\F{\dia}
\newcommand\opH{\boxM}
\newcommand\opP{\diaM}

\newcommand\Y{\Previous}

\newcommand\fildia[1]{\opP}
\newcommand\filcir[1]{\Y}
\newcommand\filboc[1]{\opH}

\newcommand\true{\emm{True}}
\newcommand\false{\emm{false}}

%% file: macros.tex
\newcommand{\stl}{STL}
\newcommand{\iostl}{IA-STL}

\newcommand{\f}{\varphi}

\newcommand{\R}{\mathbb{R}}
\newcommand{\N}{\mathbb{N}}
\newcommand{\T}{\mathbb{T}}

\newcommand{\metric}{\mathcal{M}}

\DeclareMathOperator\sign{sign}

\renewcommand{\true}{\textsf{true}}
\renewcommand{\false}{\textsf{false}}

\newcommand{\predicate}{\psi}
\newcommand{\predicates}{\Psi}

\newcommand{\aut}{\mathcal{A}}

\newcommand{\states}{Q}
\newcommand{\init}{I}
\newcommand{\final}{F}
\newcommand{\transitions}{\Delta}

\newcommand{\ppath}{\pi}
\newcommand{\paths}{\Pi}

\newcommand{\game}{\mathcal{G}}

\newcommand{\targets}{W}
\newcommand{\insF}{\operatorname{f}}
\newcommand{\insC}{\operatorname{c}}
\newcommand{\preF}{\operatorname{pre_{force}}}
\newcommand{\preC}{\operatorname{pre_{coop}}}
\newcommand{\sregion}{\operatorname{r}}

\newcommand{\strategy}{\sigma}

\newcommand{\Force}{\mathrm{Force}}
\newcommand{\Coop}{\mathrm{Coop}}
\newcommand{\strategyTrans}{\Delta_{S}}

\newcommand{\visited}{R}
\newcommand{\inprefix}{\tau}
\newcommand{\outprefix}{\omega}
\newcommand{\incurrent}{\textsf{in}}

\newcommand{\valcurrent}{\textsf{val}}
\newcommand{\simulate}{\textsf{simulate}}
\newcommand{\last}{\textsf{last}}
\newcommand{\flag}{b}
\newcommand{\pso}{\textsf{pso}}
\newcommand{\strat}{\mathcal{S}}
\newcommand{\createstrat}{\textsf{new}_{\strat}}
\newcommand{\explorestrat}{\textsf{explore}_{\strat}}
\newcommand{\adaptivetesting}{\textsf{adaptive\_testing}}
\newcommand{\initstate}{\textsf{init}}
\newcommand{\budget}{N}

%% file: intro.tex

\section{Introduction}
\label{sec:intro}

Cyber-physical systems (CPS) are becoming ubiquitous in many aspects of our lives. CPS applications 
combine computational and physical components and operate in sophisticated and unpredictable environments.
With the recent rise of machine learning, 
CPS are becoming more and more complex and ensuring their {\em safe operation} is an extremely challenging 
task.  Despite tremendous progress in the past decade, formal verification of CPS still suffers from 
scalability issues and is not an option for analysing realistic systems of high size and complexity.

The onerous exhaustive verification of CPS designs is in practice often replaced by more pragmatic {\em simulation-based testing}~\cite{BartocciDDFMNS18}. This a-priori ad-hoc 
activity can be made more systematic and rigorous by enriching automated test generation with formal specifications.  
{\em Signal Temporal Logic} (STL)~\cite{MalerN13} is a popular specification language for expressing properties of CPS. STL admits {\em robustness} semantics~\cite{fainekos-robust} 
that allows measuring how far is an observed behavior from violating a specification. 
{\em Falsification-based testing}~\cite{staliro,avstl} is a method that uses robustness evaluation to guide the system-under-test (SUT) 
to the specification violation. This successful testing approach provides effective means to detect bugs in CPS designs, but in case that 
no violation witness is detected, this method gives little information  
about design correctness. In addition, falsification testing typically stops upon detection of the first violation, thus reporting at most one fault 
at a time.
The confidence in the design correctness can be achieved by introducing a notion of {\em coverage} to the testing activity -- 
the design is considered to be correct if it passes all tests in a test suite that covers a sufficient number and variety of tests according to the 
chosen coverage metric.

\begin{figure*}[!ht]
\centering
\resizebox{0.6\width}{0.6\height}{ 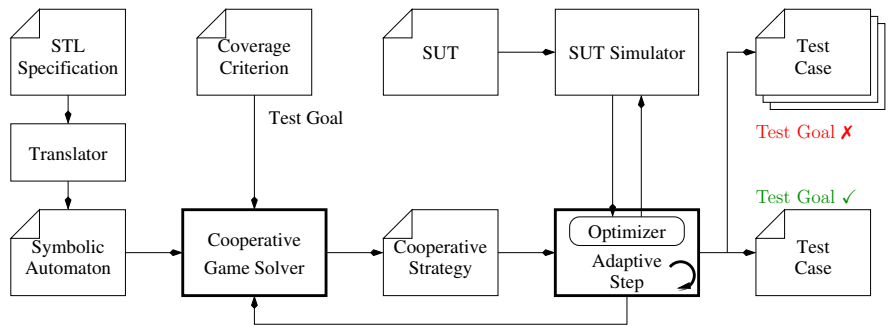 }
\caption{Adaptive testing - overview.}
\label{fig:overview}
\end{figure*}

We propose a novel adaptive test methodology for generating tests with {\em specification coverage}~\cite{TanSL04}. 
Intuitively, a test suite covers 
a specification if each requirement formalized in the specification is exercised by at least one test in the suite in a meaningful way.
The overview of our approach is shown in Figure~\ref{fig:overview}. We start with system requirements formalized in a variant of 
STL~\cite{FerrereNDIK19} that distinguishes input variables 
(controlled by the tester) from output variables (controlled by the SUT). We translate a STL specification to an equivalent 
symbolic automaton. We then define coverage on the symbolic automaton as well as the test goals needed to achieve this coverage.
To reach a test goal, we formulate a {\em cooperative reachability game} that is played between the tester and 
the system. The goal of the game is for the tester to bring the SUT to a given state. The principle of cooperative games is 
that the two players are not necessarily adversarial to each other. The state-space of the game is partitioned into 
three zones: a {\em safe} zone in which we can advance the game without cooperation from the opponent, 
a {\em no-hope} zone from which we cannot advance the game regardless of the opponent moves, and a {\em possibly winning} zone 
from which we can advance the game if the opponent is willing to cooperate. The outcome of this game is a {\em cooperative strategy}. 
In order to execute this strategy when in a possibly winning zone, 
the tester needs to steer the SUT to select a move that allows advancing the game. We call this part of the procedure an 
{\em adaptive step} and formulate the problem of 
guiding the cooperative SUT move as an optimization problem in which we use particle swarm optimization (PSO)~\cite{pso1995} to efficiently 
direct the SUT behavior in a desired direction. The outcome of this activity is a set of test cases, one of which reaches the test goal. 
We use these test cases to measure coverage and define the next test goal. We repeat this process until we reach the desired level 
of coverage or until we have used the maximum number of simulations. 
We implement our approach in a prototype tool and evaluate it on several examples, including a case study involving an aircraft elevator control system.  We demonstrate the effectiveness of adaptive testing to 
efficiently generate test cases that reach test objectives and increase specification coverage.

We summarize the main contribution of the paper:
\begin{itemize}
\item {\bf Specification coverage:} we propose a new notion of coverage for STL specifications.
\item {\bf Testing methodology:} we develop a novel adaptive testing methodology that combines cooperative games with numerical optimization:
\begin{itemize}
\item {\bf Test case generation:} incremental generation of individual test cases that achieve a given objective, and
\item {\bf Test suite generation:} steering of system executions that systematically increases specification coverage.
\end{itemize}  
\item {\bf Implementation:} we implement the proposed method in a prototype tool and evaluate it on a case study from the avionics domain.
\end{itemize}

%
%
%

%% file: overview.pdf_t
\begin{picture}(0,0)%
\includegraphics{overview.pdf}%
\end{picture}%
\setlength{\unitlength}{3947sp}%
\begingroup\makeatletter\ifx\SetFigFont\undefined%
\gdef\SetFigFont#1#2#3#4#5{%
  \reset@font\fontsize{#1}{#2pt}%
  \fontfamily{#3}\fontseries{#4}\fontshape{#5}%
  \selectfont}%
\fi\endgroup%
\begin{picture}(9174,3324)(889,-4573)
\put(8701,-2611){\makebox(0,0)[lb]{\smash{{\SetFigFont{12}{14.4}{\rmdefault}{\mddefault}{\updefault}{\color[rgb]{1,0,0}Test Goal \ding{55} }%
}}}}
\put(8701,-3286){\makebox(0,0)[lb]{\smash{{\SetFigFont{12}{14.4}{\rmdefault}{\mddefault}{\updefault}{\color[rgb]{0,.56,0}Test Goal $\checkmark$ }%
}}}}
\end{picture}%

%% file: related.tex

\section{Related Work}
\label{sec:related}


\noindent \textbf{Fault-based testing.} Fault-based testing~\cite{JiaH11} consists in introducing a fault in 
a system implementation and then computing the test that can detect/reject it. A typical example
is mutation testing where the modification of the original implementation is called \emph{mutant}. The tests causing 
a different behavior of the mutant with respect to the original implementation are said to \emph{kill the mutant}. The 
coverage of the test suite is measured by the percentage of mutants that can be killed. Our approach 
is complementary to fault-based testing. However, our notion of coverage measures the percentage
of the states/transitions visited (during testing) of a symbolic automata generated from a 
Signal Temporal Logic (STL)~\cite{MalerN13} requirement with input/output signature. 

\noindent \textbf{Falsification-based testing.} Falsification-based testing (see~\cite{BartocciDDFMNS18} for a survey) is a well-established technique 
to generate tests  violating requirements for CPSS models. This approach consists
in exercising the model with different input sequences and by monitoring each simulation trace  with 
respect to an STL~\cite{MalerN13} requirement. The use of STL quantitative semantics~\cite{fainekos-robust}  
is key to provide an indication as how far the trace is from violating the requirement and it 
is used as a fitness function for meta-heuristic optimization algorithms to guide the search of the input sequences violating the requirement.  

Although falsification-based testing has proven 
to be effective in many practical applications, the focus of this approach is solely on finding a bug and 
hence does not attempt to increase coverage. It 
also remains in  general agnostic to the syntactic 
and semantic structure of STL specification.  Our approach exploits instead the structure of the symbolic 
automaton generated from the STL requirement that we want to test and it develops a strategy to generate 
tests that increase the specification coverage.

 
 \noindent  \textbf{Model-based testing.}  In  model-based testing~\cite{FellnerKSTW17,AichernigLN13,AichernigBJKST15,Tretmans08,Dokhanchi2015}
 a model of the desired behavior of the system under test is exploited to derive the testing strategies. 
 The coverage is measured in terms of the percentage of the  model's components visited during testing. 
 Our approach belongs to this class of testing methods, where the model is the symbolic automata 
 generated from an STL requirement and the test cases are generated as winning strategies of 
 a cooperative reachability game.

\noindent \textbf{Adapting testing.} Adaptive testing~\cite{BloemFGKPRR19,BloemHRS15,AichernigBJKST15}
require the test cases to adapt with respect to the  SUT behavior observed at runtime in order to achieve the goal.
This is particularly important when the SUT are reactive systems interacting with their environment.
Our approach is also adaptive (see Figure~\ref{fig:overview}), because it requires to simulate and monitor the SUT in order 
to optimize the input that tester should provide at each time step to find the winning strategy.

\noindent \textbf{Testing as a game.} As Yannakakis formulated first  in~\cite{Yannakakis2004}, 
testing can be seen as game between a tester that aims to find the inputs revealing the faults in the system
under test (SUT) and the SUT producing outputs while hiding the internal behavior. In~\cite{DavidLLN08} the 
authors  presents a game-theoretic approach to the
testing of real-time systems. They model 
the systems as Timed I/O Game Automata and they specify
the test purposes as Timed CTL formulas. Using the timed game solver UPPAAL-TIGA they are 
able to synthesize winning strategies used to conduct black-box conformance testing of the systems. 
While in the context of timed automata, the game-theoretic approach is well-explored (see also the work of~\cite{HenryJM18}), 
to the best of our knowledge our approach is the first that can be applied directly to
MATLAB\texttrademark\; Simulink/Stafeflow models. 


\noindent  \textbf{Property-based coverage.} In~\cite{TanSL04}, the authors introduce the notion of \emph{property-based 
coverage metric} for linear temporal logic (LTL) specifications. Their approach operates on the syntax tree
of the LTL specification. The metric, based on requirement 
mutation, measures how well a property has been tested by a test-suite by 
checking the subformulae of the LTL requirement covered by a set of tests.
The main differences with our approach is the specification language (we use STL instead of LTL)
and the use of the symbolic automata generated from the specification 
both for measuring the coverage and for generating tests using cooperative games.

%% file: background.tex

\section{Background}
\label{sec:background}


\subsection{Signals and Systems}
\label{sec:signals}

Let $X = \{ x_{1}, \ldots, x_{m} \}$ be a finite set of real-valued variables. 
A valuation $v~:~X \rightarrow \R$ For $x \in X$ maps a variable $x \in X$ to a real value.
Given two disjoint sets of variables $X_{1}$ and $X_{2}$ and their associated 
valuation mappings $v_{1}~:~X_{1} \rightarrow \R$ and 
$v_{2}~:~X_{2} \rightarrow \R$, we denote by $v = v_{1}~||~v_{2}$ {\em 
valuation composition} $v~:~(X_{1} \cup X_{2}) \rightarrow \R$ 
such that $v(x) = v_{1}(x)$ if $x \in X_{1}$, and $v(x) = v_{2}(x)$ otherwise.
A {\em signal} $\textbf{v}$   is a sequence $v_{1}, v_{2},  \ldots, v_{n}$
of valuations\footnote{Given $x \in X$, we will abuse notation and denote by $x_i$ the valuation $v_i(x)$ projected to $x$ whenever it 
is clear from the context.} over $X$. We denote by $|\textbf{v}| = n$ the {\em length} of signal $\textbf{v}$.

Let $X_{I}$ be a set of {\em input} and $X_{O}$ a set of {\em output} variables. 
We consider non-linear discrete-time systems and assume that such a system $S$  is given in the form of a set of 
difference equations.


\begin{example}
\label{ex:sut}
Let $a,b \in X_I$ be input variables and $c,d \in X_{O}$ output variables. We define two simple stateless systems 
$S_{1}$ and $S_{2}$ over $X_{I} \cup X_{O}$:
\begin{equation}
\small{\begin{array}{llclllcl}
S_{1}: & c(t) & = & a(t)                 & S_{2}: & c(t) & = & 2a(t)+ b(t)      \\
           & d(t) & = & a(t) + b(t) + 2 &            & d(t) & = & a(t) +10 - b(t) \\

\end{array}}
\end{equation}
\end{example}
\vspace{-5pt}
\subsection{Interface-Aware Signal Temporal Logic}
\label{sec:stl}

We consider Signal Temporal Logic ($\stl$) with 
inputs and outputs, {\em past} and {\em future} operators, {\em quantitative} semantics and 
interpreted over discrete time. 
%
%
An {\em interface-aware signal temporal logic} (\iostl\ ) specification $\phi$ over $X$ is a tuple $(X_I, X_O, \f)$, where $X_I, X_O \subseteq X$, 
$X_I \cap X_O = \emptyset$, $X_I \cup X_O = X$ and $\f$ is an STL formula.
The syntax of a STL formula $\f$ over $X$ is defined by the following grammar:
{\small $$
\f := f(Y) \sim c~|~\neg \f~|~\f_1 \vee \f_2~|~\f_1 \until_I \f_2~|~\f_1 \since_I \f_2 
$$}
\noindent where $Y \subseteq X$, $\sim \in \{ <, \leq \}$, $c \in \R$ and $I$ is of the form $[a,b]$ or $[a, \infty)$ such that 
$a$ and $b$ are in $\N$ and $0 \leq a \leq b$. 

We equip IA-STL with standard quantitative semantics using the notion of {\em robustness}.
Let $\f$ be an STL formula and $w$ a signal trace.
We define the robustness $\rho(\f,w,t)$ by induction as follows:
\small{\begin{align*}
\rho(\true, w, t) &= +\infty \\
\rho(f(Y) > 0, w, t) &= f(w_Y[t]) \\
\rho(\lnot \f, w, t) &= -\rho(\f, w, t) \\
\rho(\f_1 \lor \f_2, w, t) &= \max \left\{ \rho(\f_1, w, t), \rho(\f_2, w, t) \right\} \\
\rho(\f_1 \until_I \f_2, w, t) &= \sup_{t' \in (t \oplus I) \cap \T} \min\left\{\begin{array}{c} \rho(\f_2, w, t') , \\ \inf_{t'' \in (t,t')} \rho(\f_1, w, t'') \end{array}\right\} \\
\rho(\f_1 \since_I \f_2, w, t) &= \sup_{t' \in (t \ominus I) \cap \T} \min\left\{\begin{array}{c} \rho(\f_2, w, t') , \\ \inf_{t'' \in (t',t)} \rho(\f_1, w, t'') \end{array}\right\} \\
\end{align*} }

\noindent where $\oplus$ and $\ominus$ are Minkowski sum and difference, $\forall a \in \R$, $\sign(a) \cdot \infty = +\infty$ if $a > 0$, $-\infty$ otherwise. 
Other Boolean/temporal operators, such as implication ($\to$), always ($\Glob_{I}$), eventually ($\F_{I}$), historically ($\opH_I$) and once ($\opP_{I}$) are 
derived from the basic operators using the standard procedure.


\begin{example}
\label{ex:spec}
Let $\phi = (\{a,b\},\{c,d\}, \varphi)$ be an IA-STL specification
defined over input variables $a$ and $b$, and output variables $c$ and $d$. 
The specification states that every time 
$a$ is greater or equal to $4$ for exactly two time units, either $b$ is negative and $c$ must be within one time unit
greater or equal to $4$, or $b$ is positive and $d$ must be greater or equal to $6$ within one time unit.

$$
{\small \begin{array}{lcll}
\varphi & = & \Glob(\opH_{[0,1]} a \geq 4 \to & ((b \leq 0 \wedge \F_{[0,1]}c \geq 4) \vee \\
&&& (b > 0 \wedge \F_{[0,1]} d \geq 6)) \\
\end{array}}
$$

It is not hard to see that $S_{1}$ satisfies $\phi$, while $S_2$ violates it. For instance, a witness of $S_{2}$ violating 
$\phi$ is an input signal in which $a$ equals to $4$ and $b$ is greater than $2$. 
\end{example}

\subsection{Symbolic Automata}
\label{sec:sa}

A metric space is a set $\metric$ possessing a distance $d$ among its elements and satisfying identity, symmetry and triangle inequality constraints. 
%
%
Given a set $M$, an element $m \in \mathcal{M}$ and $M \subseteq \mathcal{M}$, we can lift the definition of a distance to 
reason about the distance\footnote{Since $d(m,M)$ is comparing an element to a set, strictly speaking it is not a distance.} 
between an element $m$ of $\mathcal{M}$  and the subset $M$ of $\mathcal{M}$ to define a Hausdorff-like measure. In this extension, we 
extend the set of reals with the infinity $\infty$ element. We also need a special value when we compare $m$ to an empty set 
and define $d(m, \emptyset) = \infty$.
{\small $$ 
d(m,M) = \left \{
				\begin{array}{ll}
				\infty & \textrm{if }  M \textrm{ is empty} \\
				\min_{m' \in M} d(m,m') & \textrm{otherwise.}
				\end{array}
				\right.
$$}

\begin{definition}[Predicate]
\label{def:pred}
The following grammar defines the syntax of a {\em predicate} $\predicate$ over $X$:
$\predicate := \top~|~f(Y) > 0~|~\neg \psi~|~\psi_{1} \vee \psi_{2}$,  
where $Y \subseteq X$. 
\end{definition}

$\predicates(X)$ denotes all predicates over $X$. 
We lift the definition of a distance between two valuations to the distance between a valuation and a predicate. 

\begin{definition}[Valuation-predicate distance]
\label{def:pred-val}
Given a valuation $v \in V(X)$ and a predicate $\psi \in \predicates (X)$, we have that:
$
d(v, \psi) = \min_{v' \models \psi}  \max_{x \in X} ~d(v(x), v'(x)).
$
\end{definition}

We now define {\em symbolic} and {\em symbolic weighted automata}. 

\begin{definition}[Symbolic Automata]
\label{def:wsa}
A {\em symbolic automaton} (SA) $\aut$ is the tuple $\aut = (X, \states, \init, \final, \transitions)$, where $X$ is a finite set of variables partitioned 
into the disjoint sets $X_I$ of {\em input} and $X_O$ of {\em output} variables.
$\states$ is a finite set of {\em locations}, $\init \subseteq \states$ is the set of {\em initial} states, $\final \subseteq \states$ is 
the set of {\em final} states and $\transitions \subseteq \states \times \predicates(X) \times \states$ is the {\em transition relation}. 
\end{definition}

A {\em path} $\ppath$ in $\aut$ is a finite alternating sequence of locations and transitions 
$\ppath = q_{0}, \delta_{1}, q_{1}, \ldots, q_{n-1}, \delta_{n}, q_{n}$ such that $q_{0} \in \init$ and for all $1 \leq i \leq n$, $(q_{i-1}, \delta_{i}, q_{i}) \in \transitions$. 
We say that the path $\pi$ is {\em accepting} if $q_{n} \in \final$. We say that a signal $w = v_{1}, v_{2}, \ldots, v_{n}$ {\em induces} a path 
$\ppath = q_{0}, \delta_{1}, q_{1}, \ldots, q_{n-1}, \delta_{n}, q_{n}$ in $\aut$ if for all $1 \leq i \leq n$, $v_{i} \models \predicate_{i}$, where 
$\delta_{i} = (q_{i-1}, \predicate_{i}, q_{i})$. We denote by 
$\paths(w) = \{ \ppath~|~\ppath \in \final \; \textrm{and} \; w \; \textrm{induces} \; \ppath \; \textrm{in} \; \aut \}$ the 
set of all accepting paths in $\aut$ induced by trace $w$.

A SA $\aut$ is {\em deterministic} iff $I = \{q\}$ for some $q \in Q$ and for all $\delta = (q, \psi, q'), \delta' = (q, \psi', q'') \in \Delta$ such that $\delta \neq \delta'$, 
$\psi \wedge \psi'$ is unsatisfiable. A SA $\aut$ is {\em complete} iff for all $q \in Q$, $v \in V(X)$, there exists 
$(q, \psi, q') \in \Delta$ such that $v \models \psi$.  
%
Additionally we require all predicates to be satisfiable: $\forall (q,\predicate, q') \in \transitions: \exists v \in V(X): v \models \predicate$.

%

\begin{example}
\label{ex:aut}
Consider the IA-STL specification $\phi$ from Example~\ref{ex:spec}.
Figure~\ref{fig:illustrating} depicts the deterministic symbolic automaton $\mathcal{A}_{\phi}$ associated with $\phi$.

\begin{figure}[!ht]
\centering
\scalebox{0.65}{ 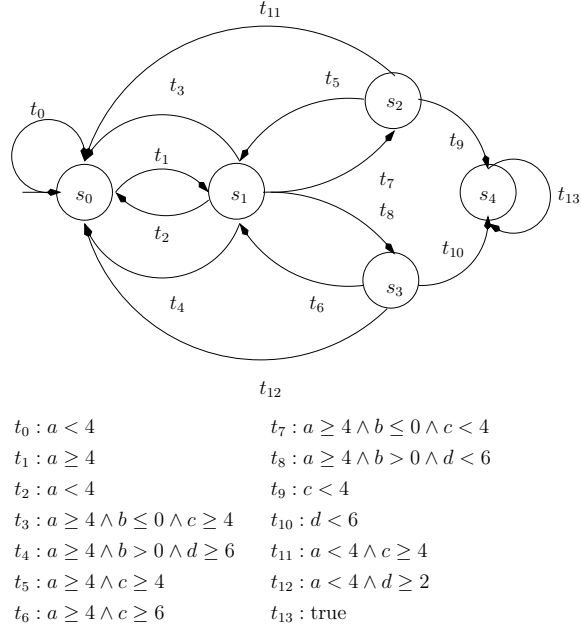 }
\caption{Symbolic automaton $\mathcal{A}_{\phi}$.}
\label{fig:illustrating}
\end{figure}
\end{example}


\subsection{Specification Coverage}
\label{sec:coverage}

A {\em test coverage criterion} $C$ defines a set of testing requirements that a {\em test suite} $T$ must fulfil. Given a 
test $\tau \in T$ and a requirement $c \in C$, we denote by $\tau \models c$ the fact that $\tau$ {\em satisfies} the criterion $c$. 
We define by $R(\tau,C) = \{ c~|~c \in C \text{ and } \tau \models c \}$ and $R(T,C) = \bigcup_{\tau \in T} R(\tau, C)$ the set of test coverage requirements from the 
criterion $C$ satisfied by the test $\tau$ and the test suite $T$, respectively. The {\em test coverage} of $T$, denoted by $K(C,T)$ is the ratio between 
the number of testing requirements in $C$ that $T$ fulfils and the total number of testing requirements defined by $C$, 
that is $K(C,T) = \frac{|R(T,C)|}{|C|}$.

Test coverage criteria are typically defined on the finite state machine that specifies the intended behavior of the 
SUT (model-based or black-box testing) or on the actual code of the SUT implementation (white-box testing). 
Test coverage metrics have been a vivid area of research. We focus in this paper on two simple 
criteria, defined on the symbolic automaton derived from the IA-STL specification -- 
{\em location} and {\em transition} coverage criteria  
that we formally define in the remainder of this section.

Let $\phi = (X_{I}, X_{O}, \varphi)$ be an IA-STL specification, $\aut_{\phi}$ its associated symbolic automaton 
and $S$ the system model. A test $\tau$ is the sequence $\textbf{v}_{I}$ of input variable valuations. The location 
coverage criterion $C_{Q} = Q$ is the set of all locations in $\aut_{\phi}$. Given a location 
$q \in C_{Q}$ and a test $\tau$, we have that $\tau$ satisfies the criterion $c$ if the input signal $\tau$ induces a path 
$\pi = q_{0}, \delta_{1}, \ldots, q_{n-1}, \delta_{n}, q_{n}$ and $q = q_i$ for some $i \in [0,n]$. 
Similarly, we define the transition coverage criterion $C_{\Delta} = \Delta$ as the set of transitions in 
$\aut_{\phi}$. 

\begin{example}
\label{ex:test}
Consider the system $S_{2}$ from Example~\ref{ex:sut}, the IA-STL specification $\phi$ from Example~\ref{ex:spec} and 
its associated symbolic automaton $\aut_{\phi}$ from Example~\ref{ex:aut}. Equation~\ref{eq:suite} shows the test suite $T = \{ \tau_{1}, \tau_{2} \}$ 
and the induced outputs $S_{2}(\tau_{1})$ and $S_{2}(\tau_{2})$. The trace $\tau_{1}~||~S_{2}(\tau_{1})$ induces the run $s_{0}, t_{0}, s_{0}, t_{1}, s_{1}, t_{2}, s_{0}$ 
in $\aut_{\phi}$, 
while the test $\tau_{2}~||~S_{2}(\tau_{2})$ induces the run $s_{0}, t_{1}, s_{1}, t_{7}, s_{2}, t_{11}, s_{0}$ in $\aut_{\phi}$.

\begin{equation}
\label{eq:suite}
\small{\begin{array}{|rl | rrr | rl | rrr|}
\hline
\tau_{1}: 	 & a: & 3 & 4 & 3 & \tau_{2}: 	 & a: & 4 & 4 & 2\\
	            & b: & 2 & 2 & 2 & & b: & 2 & -8 & 2 \\
S_{2}(\tau_{1}):	 & c: & 8 & 10 & 8 & S_{2}(\tau_{2}):	 & c: & 10 & 0 & 6 \\
		& d: & 11 & 12 & 11 &  & d: & 12 & 22 & 10 \\
\hline
\end{array}}
\end{equation}
\end{example}

In this example, we have that $R(\tau_{1}, C_{Q}) = \{ s_{0}, s_{1} \}$, $R(\tau_{1}, C_{\Delta}) = \{ t_{0}, t_{1}, t_{2}\}$, 
$R(\tau_{2}, C_{Q}) = \{ s_{0}, s_{1}, s_{2} \}$ and $R(\tau_{2}, C_{\Delta}) = \{ t_{1}, t_{7}, t_{11}\}$. 
It follows that $R(T, C_{Q}) = \{ s_{0}, s_{1}, s_{2} \}$ and $R(T, C_{\Delta}) = \{ t_{0}, t_{1}, t_{2}, t_{7}, t_{11}\}$. 
Hence, these two test cases together achieve $K(C_{Q}, T) = 60 \%$ location and $K(C_{\Delta}, T) = 36 \%$ transition 
coverage.

\subsection{Specification-based Testing as Optimization}
\label{sec:falsification}

The quantitative semantics of STL allows to formulate the testing problem of finding an input sequence that violates an IA-STL specification $\varphi$ 
as an optimization problem over the input sequence $\tau$ as follows: {\small $$ \min_{\tau} \rho(w, \varphi)  \textrm{ s.t. }  w = \tau~||~S(\tau) $$}


This testing approach is also known as {\em falsification testing} in the literature. The optimization 
procedure can take multiple forms: gradient descent, genetic algorithms, simulated annealing, etc. 

In this paper, we use the particle swarm 
optimization (PSO)~\cite{pso1995}, a randomized search algorithm in 
which a collection of points in the search space are updated at each iteration 
to move closer (on average) to a global optimal solution. This meta-heuristic 
does not require a differentiable objective function and it is easy 
to parallelize.

%% file: illustrating.pdf_t
\begin{picture}(0,0)%
\includegraphics{illustrating.pdf}%
\end{picture}%
\setlength{\unitlength}{3947sp}%
\begingroup\makeatletter\ifx\SetFigFont\undefined%
\gdef\SetFigFont#1#2#3#4#5{%
\reset@font\fontsize{#1}{#2pt}%
\fontfamily{#3}\fontseries{#4}\fontshape{#5}%
\selectfont}%
\fi\endgroup%
\begin{picture}(5303,6382)(1388,-7859)
\put(1426,-5686){\makebox(0,0)[lb]{\smash{{\SetFigFont{12}{14.4}{\rmdefault}{\mddefault}{\updefault}{\color[rgb]{0,0,0}$t_0: a < 4$}%
}}}}
\put(3901,-6886){\makebox(0,0)[lb]{\smash{{\SetFigFont{12}{14.4}{\rmdefault}{\mddefault}{\updefault}{\color[rgb]{0,0,0}$t_{11}: a < 4 \wedge c \geq 4$}%
}}}}
\put(3901,-5986){\makebox(0,0)[lb]{\smash{{\SetFigFont{12}{14.4}{\rmdefault}{\mddefault}{\updefault}{\color[rgb]{0,0,0}$t_8: a \geq 4 \wedge b > 0 \wedge d < 6$}%
}}}}
\put(3901,-6586){\makebox(0,0)[lb]{\smash{{\SetFigFont{12}{14.4}{\rmdefault}{\mddefault}{\updefault}{\color[rgb]{0,0,0}$t_{10}: d< 6$}%
}}}}
\put(3901,-7186){\makebox(0,0)[lb]{\smash{{\SetFigFont{12}{14.4}{\rmdefault}{\mddefault}{\updefault}{\color[rgb]{0,0,0}$t_{12}:a < 4 \wedge d \geq 2$}%
}}}}
\put(1426,-7486){\makebox(0,0)[lb]{\smash{{\SetFigFont{12}{14.4}{\rmdefault}{\mddefault}{\updefault}{\color[rgb]{0,0,0}$t_6: a \geq 4 \wedge c \geq 6$}%
}}}}
\put(1426,-6886){\makebox(0,0)[lb]{\smash{{\SetFigFont{12}{14.4}{\rmdefault}{\mddefault}{\updefault}{\color[rgb]{0,0,0}$t_4: a\geq 4 \wedge b > 0 \wedge d \geq 6$}%
}}}}
\put(3901,-5686){\makebox(0,0)[lb]{\smash{{\SetFigFont{12}{14.4}{\rmdefault}{\mddefault}{\updefault}{\color[rgb]{0,0,0}$t_7: a \geq 4 \wedge b \leq 0 \wedge c < 4$}%
}}}}
\put(1426,-6586){\makebox(0,0)[lb]{\smash{{\SetFigFont{12}{14.4}{\rmdefault}{\mddefault}{\updefault}{\color[rgb]{0,0,0}$t_3: a\geq 4 \wedge b \leq 0 \wedge c \geq 4$}%
}}}}
\put(1426,-7186){\makebox(0,0)[lb]{\smash{{\SetFigFont{12}{14.4}{\rmdefault}{\mddefault}{\updefault}{\color[rgb]{0,0,0}$t_5: a \geq 4 \wedge c \geq 4$}%
}}}}
\put(1426,-6286){\makebox(0,0)[lb]{\smash{{\SetFigFont{12}{14.4}{\rmdefault}{\mddefault}{\updefault}{\color[rgb]{0,0,0}$t_2: a < 4$}%
}}}}
\put(1426,-5986){\makebox(0,0)[lb]{\smash{{\SetFigFont{12}{14.4}{\rmdefault}{\mddefault}{\updefault}{\color[rgb]{0,0,0}$t_1:a \geq 4$}%
}}}}
\put(3901,-6286){\makebox(0,0)[lb]{\smash{{\SetFigFont{12}{14.4}{\rmdefault}{\mddefault}{\updefault}{\color[rgb]{0,0,0}$t_9: c < 4$}%
}}}}
\put(3901,-7486){\makebox(0,0)[lb]{\smash{{\SetFigFont{12}{14.4}{\rmdefault}{\mddefault}{\updefault}{\color[rgb]{0,0,0}$t_{13}: \text{true}$}%
}}}}
\put(1651,-2611){\makebox(0,0)[b]{\smash{{\SetFigFont{12}{14.4}{\rmdefault}{\mddefault}{\updefault}{\color[rgb]{0,0,0}$t_0$}%
}}}}
\put(3901,-5311){\makebox(0,0)[b]{\smash{{\SetFigFont{12}{14.4}{\rmdefault}{\mddefault}{\updefault}{\color[rgb]{0,0,0}$t_{12}$}%
}}}}
\put(3901,-1636){\makebox(0,0)[b]{\smash{{\SetFigFont{12}{14.4}{\rmdefault}{\mddefault}{\updefault}{\color[rgb]{0,0,0}$t_{11}$}%
}}}}
\put(5776,-2911){\makebox(0,0)[rb]{\smash{{\SetFigFont{12}{14.4}{\rmdefault}{\mddefault}{\updefault}{\color[rgb]{0,0,0}$t_9$}%
}}}}
\put(5776,-3961){\makebox(0,0)[rb]{\smash{{\SetFigFont{12}{14.4}{\rmdefault}{\mddefault}{\updefault}{\color[rgb]{0,0,0}$t_{10}$}%
}}}}
\put(6676,-3436){\makebox(0,0)[lb]{\smash{{\SetFigFont{12}{14.4}{\rmdefault}{\mddefault}{\updefault}{\color[rgb]{0,0,0}$t_{13}$}%
}}}}
\put(5026,-3586){\makebox(0,0)[b]{\smash{{\SetFigFont{12}{14.4}{\rmdefault}{\mddefault}{\updefault}{\color[rgb]{0,0,0}$t_8$}%
}}}}
\put(5026,-3286){\makebox(0,0)[b]{\smash{{\SetFigFont{12}{14.4}{\rmdefault}{\mddefault}{\updefault}{\color[rgb]{0,0,0}$t_7$}%
}}}}
\put(2851,-3811){\makebox(0,0)[b]{\smash{{\SetFigFont{12}{14.4}{\rmdefault}{\mddefault}{\updefault}{\color[rgb]{0,0,0}$t_2$}%
}}}}
\put(2851,-3061){\makebox(0,0)[b]{\smash{{\SetFigFont{12}{14.4}{\rmdefault}{\mddefault}{\updefault}{\color[rgb]{0,0,0}$t_1$}%
}}}}
\put(4351,-4486){\makebox(0,0)[b]{\smash{{\SetFigFont{12}{14.4}{\rmdefault}{\mddefault}{\updefault}{\color[rgb]{0,0,0}$t_6$}%
}}}}
\put(4501,-2311){\makebox(0,0)[b]{\smash{{\SetFigFont{12}{14.4}{\rmdefault}{\mddefault}{\updefault}{\color[rgb]{0,0,0}$t_5$}%
}}}}
\put(3001,-4486){\makebox(0,0)[b]{\smash{{\SetFigFont{12}{14.4}{\rmdefault}{\mddefault}{\updefault}{\color[rgb]{0,0,0}$t_4$}%
}}}}
\put(3001,-2386){\makebox(0,0)[b]{\smash{{\SetFigFont{12}{14.4}{\rmdefault}{\mddefault}{\updefault}{\color[rgb]{0,0,0}$t_3$}%
}}}}
\put(2101,-3436){\makebox(0,0)[b]{\smash{{\SetFigFont{12}{14.4}{\rmdefault}{\mddefault}{\updefault}{\color[rgb]{0,0,0}$s_0$}%
}}}}
\put(3601,-3436){\makebox(0,0)[b]{\smash{{\SetFigFont{12}{14.4}{\rmdefault}{\mddefault}{\updefault}{\color[rgb]{0,0,0}$s_1$}%
}}}}
\put(5101,-4336){\makebox(0,0)[b]{\smash{{\SetFigFont{12}{14.4}{\rmdefault}{\mddefault}{\updefault}{\color[rgb]{0,0,0}$s_3$}%
}}}}
\put(5101,-2536){\makebox(0,0)[b]{\smash{{\SetFigFont{12}{14.4}{\rmdefault}{\mddefault}{\updefault}{\color[rgb]{0,0,0}$s_2$}%
}}}}
\put(6001,-3436){\makebox(0,0)[b]{\smash{{\SetFigFont{12}{14.4}{\rmdefault}{\mddefault}{\updefault}{\color[rgb]{0,0,0}$s_4$}%
}}}}
\end{picture}%

%% file: games.tex
\section{Adaptive Testing for Specification Coverage}
\label{sec:atest}

In this section, we present the adaptive testing algorithm for specification coverage. We start by introducing 
{\em cooperative reachability games} in Section~\ref{sec:games} and then present the adaptive testing 
algorithm that combines cooperative games with numerical optimization in Section~\ref{sec:sbt}.

\subsection{Cooperative Reachability Games}
\label{sec:games}
Cooperative reachability games have been used for testing by David~et~al.~\cite{DavidLLN08}. We define our own version 
that allows cooperation at any point (not only at the beginning) and handles real valued signals.

A {\em symbolic reachability game} $\game$ is the pair $(\aut, \targets)$ where $\aut$ is a deterministic symbolic automaton and $\targets \subseteq \states$ is a set of target locations. The game is played by two players, the tester and the system. In every location $q \in \states$ starting from the initial location $q_0$ the tester moves first by picking inputs $v_I \in V(X_I)$ and in response the system picks outputs $v_O \in V(X_O)$.
An interaction in which both players take a move by choosing their values is called a turn.
A game is won by the tester if at some point a target state $q_n \in \targets$ is reached.

A test strategy $\strategy : \states^* \rightarrow 2^{V(X_I)}$ is used to determine the values the tester picks in each move, based on the trace of previously visited locations.
We generalize strategies to not only provide a single input for each move, but instead a set of possible inputs.
In the following we will focus on positional strategies $\strategy : \states \rightarrow 2^{V(X_I)}$ where the next move only depends on the current location $q$.
A strategy is called a winning strategy if a tester picking inputs according to the strategy is guaranteed to achieve the winning condition, independent of the systems choices.

The tester cannot expect to reach all location of the specification, if the system is adversarial.
Instead, we assume a setting where the system may choose outputs that help the tester.
We first give an informal overview and provide the formal definitions afterwards.

A strategy distinguishes three kinds of locations with respect to reaching the target set.
A {\em force} location is one where the tester can pick inputs such that it is guaranteed to get to a location closer to the target set;
a {\em cooperative} location is one where the tester can pick inputs such that it gets to a location closer to the target set iff the system chooses helpful outputs;
a {\em no-hope} location is any one from where the target set can never be reached.

A {\em cooperative winning strategy} is a strategy where a tester following it will always reach the target set if the system helps in all cooperative locations. We now give the formal definitions of a cooperative winning strategy and how it can be obtained from a game automaton.



To calculate a cooperative winning strategy for a game $\game$ we first define the functions $\insF : \states \times 2^{\states} \rightarrow 2^{V(X_I)}$ and $\insC : \states \times 2^{\states} \rightarrow 2^{V(X_I)}$.
The function $\insF$ takes a location $q$ and a set of locations $S$ and evaluates to the set of inputs that allow the tester to reach a location in $S$ from $q$ in a single turn, for all possible outputs of the system. Formally, we define
$\insF(q, S) = \{v_I \in V(X_I) \mid \forall v_O \in V(X_O): \exists (q, \predicate, q') \in \transitions: q' \in S \wedge v_I || v_O \models \predicate\}$.
The function $\insC$ is defined analogously, but assumes the system cooperates by existentially quantifying the outputs; 
$\insC(q, S) = \{v_I \in V(X_I) \mid \exists v_O \in V(X_O): \exists (q, \predicate, q') \in \transitions: q' \in S \wedge v_I || v_O \models \predicate\}$.

Using $\insF$ and $\insC$ we define the preimages of force and cooperative moves $\preF(S) = \{ q \mid \insF(q,S) \neq \emptyset \}$ and $\preC(S) = \{ q \mid \insC(q,S) \neq \emptyset \}$.
We use two nested fixpoints to calculate the {\em winning region}, the set of all locations for which $\targets$ can be reached.
The strategy should contain the least amount of cooperative steps possible. Therefore we only grow it by a cooperative move if we cannot grow it with a force move. The intermediate regions are sets indexed by two natural numbers, the first one counting the cooperative steps and the second one counting the force steps. We initialize $Y_{0,0} = \targets$. We extend a region by force moves such that $Y_{i,j+1} = Y_{i,j} \cup \preF(Y_{i,j})$, till we reach a fixpoint $Y_{i,\infty} =  \bigcup_j Y_{i,j}$. In that case we extend the set by a single cooperative move $Y_{i+1,0} = Y_{i,\infty} \cup \preC(Y_{i,\infty})$ and iterate these two steps until the fixpoint $Y_{\infty,\infty} = \bigcup_i Y_{i,\infty}$ is found.
This process converges, as the number of states is finite.
%
There exists a cooperative winning strategy iff the initial state $q_0$ is in the winning region $Y_{\infty,\infty}$.

A strategy can be extracted from this fixpoint calculation as follows.
Let $\sregion$ be a function mapping locations in the winning region to pairs $(i,j)$ of positive integers identifying the first (smallest) region $Y_{i,j}$ containing the location. 
Formally, $\sregion(q) = (i,j)$ such that $q \in Y_{i,j}$, but $q \not\in Y_{i',j'}$ for any $(i',j')$ lexicographically smaller than $(i,j)$ where $Y_{i,j}$ are sets from the fixpoint computation.

Let $\Force = \{q \mid \sregion(q) = (i,j+1)  \}$ and $\Coop = \{q \mid \sregion(q) = (i+1,0)  \}$ be the sets of force and cooperative locations.
Using these, the strategy function is defined as:
{\small \begin{align*}
&\strategy(q) = \begin{cases}
  \emptyset & \text{if $q \not\in Y_{\infty,\infty}$}\\
  V(X_I) & \text{if $q \in Y_{0,0}$}\\
  \insF(q, Y_{i,j}) & \text{if $q \in \Force$ and $\sregion(q) = (i,j+1)$}\\
  \insC(q, Y_{i,\infty}) & \text{if $q \in \Coop$ and $\sregion(q) = (i+1,0)$}.
\end{cases}
\end{align*}}
The good transitions according to the strategy, that is those pointing to the next smaller region either by force or cooperation, are given as:
{\small \begin{align*}
\strategyTrans = &\{ (q,\predicate,q') \in \transitions \mid q \in \Force \wedge \sregion(q) = (i,j+1) \wedge q' \in Y_{i,j}\} \cup \\
&\{ (q,\predicate,q') \in \transitions \mid q \in \Coop \wedge \sregion(q) = (i+1,0) \wedge q' \in Y_{i,\infty}\}.
\end{align*}}
The strategy automaton $\strat$ for a game $\game$ is the tuple $\strat = (X, Y_{\infty,\infty},\\ \text{init}, \strategyTrans, \Force, \Coop, \targets, \strategy)$. It is a subgraph of the automaton $\aut$ which only contains the states from the winning region $Y_{\infty,\infty}$ and the transitions defined as $\strategyTrans$. There is unique initial state $\text{init}$. The states are partitioned into $\Force$, $\Coop$ and target states $\targets$. Additionally, $\strategy$ labels each state with the set of good inputs.

\begin{example}
Consider $\mathcal{A}_{\phi}$ from Example~\ref{ex:aut}.
The cooperative winning strategy for reaching the target state $s4$ is depicted as the strategy automaton $\mathcal{S}_{\phi}$ in Figure~\ref{fig:game}. The input set given by $\strategy$ is shown symbolically as a predicate on each state. Cooperative states are bold whereas the target state is dashed. The transition predicates are the same as in Figure~\ref{fig:illustrating}.
Additionally, the regions of the fixpoint computation used to obtain the strategy are shown.

\begin{figure}[tb]
\centering
\includegraphics[width=.5\textwidth]{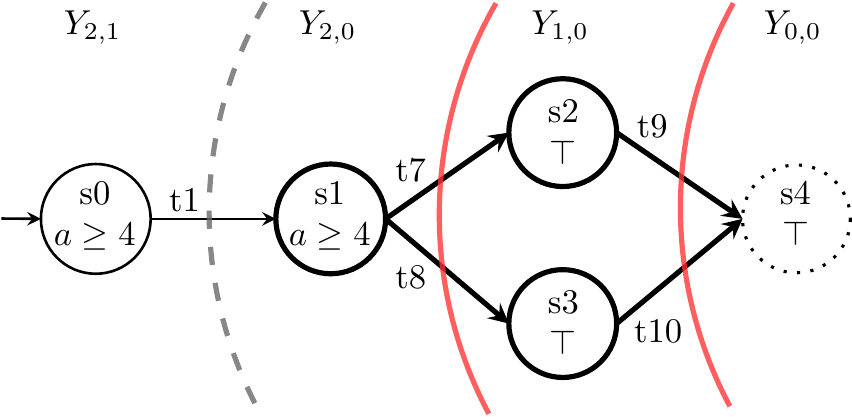}
\caption{Strategy $\mathcal{S}_{\phi}$ for target state $s4$.}
\label{fig:game}
\end{figure}
\end{example}

\subsection{Using Cooperative Games and Search-based Testing}   
\label{sec:sbt}

In Section~\ref{sec:games}, we presented a game between a tester and a system and developed a procedure for computing a cooperative strategy 
from this game that requires cooperation between the two players in order to reach the game objective. In this section, 
we propose a procedure based on PSO that facilitates the cooperation between the two players. We generalize this 
procedure to devise a method for adaptive testing with specification coverage\footnote{We assume location coverage when describing the procedure. 
However, the procedure can be adapted to transition coverage.}. 

The main procedure $\adaptivetesting$ is described in Algorithm~\ref{alg:coop}. The procedure takes as inputs: (1) the automaton $\aut$ obtained from the 
specification $\phi$, (2) the system model $S$, and (3) the user-defined budget $\budget$ that defines the maximum number of system model 
simulations. The procedures computes a test suite $T$ and the set $R(C_{Q}, T)$ of test coverage requirements from the location criterion 
$C_{Q}$ satisfied by $T$.

\begin{algorithm}[tb]
\KwIn{Specification automaton $\aut$, system $S$, budget $\budget$}
\KwOut{Test suite $T$, visited locations $R(C_{Q}, T)$}
$\visited := \emptyset$; $T := \emptyset$\;
\While{$\visited \subset Q \wedge \budget \geq 0$}{
    $\hat{q} :\in Q\backslash \visited; \aut' := \aut; \flag := \false$\;
    \While{$\neg \flag \wedge \budget \geq 0$}{
        $\strat := \createstrat(\aut', \{\hat{q}\})$\;
        $q = \initstate(\strat)$\;
        $\flag:= \explorestrat(\aut', T, \visited, N, S, q, \hat{q}, [])$\;
    }
}
\Return $(T, \visited)$\;
\caption{Algorithm $\adaptivetesting$}
\label{alg:coop}
\end{algorithm}

The procedure $\adaptivetesting$ maintains a set of generated input sequences $T$ and a set of visited states $\visited$ that are both initialized 
to empty sets (line $1$). The main loop (lines $2-9$) consists in generating tests that improve location coverage of $\aut$. The loop 
executes while there are not visited locations in the automaton and there is sufficient budget for more simulations. 
In every loop iteration, the procedure first selects a target location 
$\hat{q}$ that has not been visited yet and makes a copy of 
the input automaton $\aut$ into $\aut'$ (line $3$). In the next step, the algorithm attempts to generate a test that allows reaching the 
target location. The procedure generates a cooperative reachability strategy (see Section~\ref{sec:games}) $\strat$ from $\aut$ and $\hat{q}$ (line $5$). 
The initial location of the strategy automaton is copied to $q$ (line $6$). The cooperative reachability strategy $\strat$ is explored by the procedure  
$\explorestrat$. This procedure aims at generating an input sequence, that when executed on the system model $S$, induces a run in $\strat$ that 
reaches $\hat{q}$ from the initial location\footnote{We assume that $\explorestrat$ passes parameters by reference. In particular, the procedure 
updates $\aut'$, $T$, $\visited$ and $N$.}. If successful, the procedure returns $\true$. Otherwise, it returns $\false$ and updates the automaton 
$\aut'$ automaton $\aut'$ by removing the location that could not be reached and that is used to generate a new cooperative strategy. In both 
cases, the procedure updates the set $T$ with input sequences used to simulate the system model $S$, the set $R$ with the set of locations 
that were visited by runs induced by these input sequences and the remaining budget $\budget$.
Algorithm~\ref{alg:sbt} describes the details of $\explorestrat$. 

\begin{algorithm}[tb]
\KwIn{Automaton $\aut$, input sequences $T$, visited locations $\visited$, budget $\budget$, system $S$, initial location $q$, input prefix $\inprefix$}
\KwOut{Flag $\flag$}
$\visited := \visited \cup \{ q \}$\;
\If{$q = \hat{q}$}{
    $\Return$ $\true$\;
}
\If{$q \in \textrm{Force}$}{
    $\incurrent :\in \strategy(q)$\;
    $\inprefix := [\inprefix, \incurrent]$; $T := T \cup \{ \inprefix \}$\;
    $\outprefix := \simulate(S, \inprefix)$\;
    $\budget := \budget - 1$\;
    $\valcurrent := \last(\inprefix)~||~\last(\outprefix)$\;
    $q' :\in \{s~|~(q,\psi, s) \in \Delta_S \text{ and } \valcurrent \models \psi \}$\;
    $\Return$ $\explorestrat(\aut', T, \visited, \budget, S, q', \hat{q}, \inprefix)$\;
}\Else{
    \For{$(q, \psi, q') \in \Delta_S$}{
        $(\incurrent, \flag) := \pso(\psi, S, \budget)$\;
        \If{$\flag$}{
            $\inprefix := [\inprefix, \incurrent]; T := T \cup \{ \inprefix \}; \visited := \visited \cup \{ q' \}$\;
            $\Return$ $\explorestrat(\aut', T, \visited, \budget, S, q', \hat{q}, \inprefix)$\;
        }\Else{
           $\aut' := (X, Q, I, F, \Delta \backslash \{(q, \psi, q')\})$\;
           $\Return$ $\false$\;
       }
    }
}
\caption{Algorithm $\explorestrat$}
\label{alg:sbt}
\end{algorithm}

The procedure \emph{explore} aims at executing a cooperative strategy by using PSO to facilitate cooperation between the tester and the system.
This procedure recursively computes the test input sequence $\tau$ that induces a run from an initial location $q$ to the target location $\hat{q}$. 
It first adds the input location $q$ into the set of 
visited states (line $1$). It then checks if $q$ is the target state, and if yes, it returns $\true$. 
Otherwise, the procedure checks if the location $q$ is of type Force or Cooperative. 

If the location $q$ is a forced location (lines $5-13$), 
the algorithm first finds an input that satisfies the input predicate associated to $q$ (line $6$) by invoking an 
SMT solver and finding a model of the formula representing the input predicate. This input is appended to the existing input sequence and is added to 
the test suite $T$ (line $7$). Next, the 
system model $S$ is simulated with $\tau$ (line $8$) and the remaining simulation budget is decremented by one (line $9$). 
The procedure then picks the target location $q'$ of the transition whose predicate is satisfied by the 
computed input/output valuation from the simulation (line $11$), and the procedure reinvokes itself, but with $q'$ as the new input location and with the updated 
other parameters. 

If the location $q$ is Coop (lines $14-26$), then the algorithm repeats an exploration step for each of its outgoing 
transitions (lines $15-25$). The procedure tries to find a valuation that satisfies the predicate $\psi$ that decorates the transition. It does so by invoking 
the particle swarm optimization (PSO) procedure. We deviate from the standard PSO by adding two adaptations 
to the procedure: (1) it stops as soon as it finds a valuation that satisfies the predicate $\psi$ and enables the transition, and (2) after each simulation, it 
decrements the budget variable $\budget$ by one and terminates if the entire budget is consumed. 
Every step of the PSO search invokes a simulation of the system model with the input chosen by the PSO algorithm. 
If the search for the valuation that enables the transition is successful (lines $17-22$), then the procedure reinvokes itself with the target location $q'$ of 
the transition. Otherwise, the procedure updates the automaton $\aut'$ by removing the transition $(q, psi, q')$ and returns $\false$ (lines $21-24$).

%% file: casestudy.tex
\section{Evaluation and Experiments}
\label{sec:experimental}


\subsection{Evaluation using an Illustrative Example}
\label{sec:eva}

In this section, we illustrate qualitative outcomes of adaptive testing applied to the systems $S_{1}$ and $S_{2}$ from Example~\ref{ex:sut} with 
the formal specification $\varphi$ of its requirements from Example~\ref{ex:spec}. 

\begin{figure}[tb]
\centering
\scalebox{0.65}{ 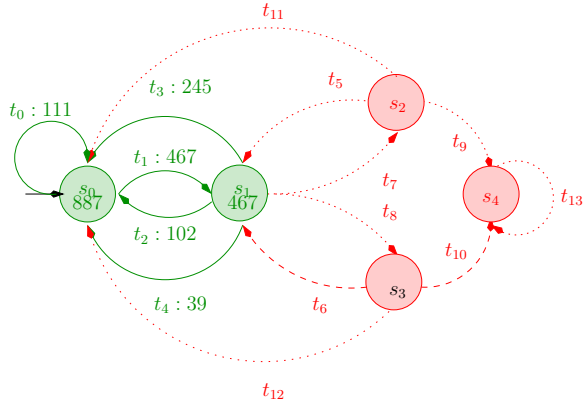 }
\caption{Specification coverage for $S_{1}$ and $\varphi$.}
\label{fig:eval1}
\end{figure}

We first apply adaptive testing to $S_{1}$. 
We do not set an 
upper bound on budget and we initialize the PSO algorithm with the maximum swarm size of $100$ and the maximum number of iterations of $100$. 
Adaptive testing procedure conducted $390$ simulations in $57s$ (including $10s$ for initializing MATLAB Simulink).
The results of this experiment are shown in 
Figure~\ref{fig:eval1}. This figure depicts the specification coverage for $S_{1}$ and $\varphi$. 
Visited locations and transitions are shown in green. We used red color to mark locations and transitions that could not be 
reached. every visited location and transition is labeled by the number of times it was visited. 

We can make the following observations. First, we could achieve $40\%$ location and  $38\%$ 
transition coverage. While this coverage may seem low, it actually cannot be improved for 
two reasons: (1)  $S_{1}$ satisfies $\varphi$, hence the error location $s_{4}$ cannot be reached 
and its incoming transitions $t_{9}$, $t_{10}$ and $t_{13}$ of $s_{4}$ cannot be enabled by any input/output 
combination allowed by the dynamics of $S_{1}$; and (2) $S_{1}$ implements only a subset of $\varphi$ (implementation choice).
For instance, $S_{1}$ always immediately satisfies the obligation $\F_{[0,1]} c\geq4$ in $\varphi$ even though the specification 
allows satisfying it with one-step delay. 
As a consequence, adaptive testing does not only gives us confidence that 
$S_{1}$ satisfies $\varphi$, it also indicates which implementation choices were made. 
In particular, we can observe that 
the green locations and transitions from Figure~\ref{fig:eval1} corresponding to the part of $\varphi$ implemented by $S_{1}$ corresponds to the specification
{\small $$
\begin{array}{l}
\Glob(\opH_{[0,1]} a \geq 4 \to  ((b \leq 0 \wedge c \geq 4) \vee  (b > 0 \wedge d \geq 6)),
\end{array}
$$}
\noindent which effectively refines $\varphi$. 

Next, we apply adaptive testing to $S_{2}$. We keep the same parameters, except that we replace location with 
with transition coverage. Adaptive testing procedure conducted 
$133$ simulations in $18s$. 
%

The main observation about this experiment is that we are able to reach all the locations in the specification, including the error location $s_{4}$ 
that we entered via both transitions $t_{9}$ and $t_{10}$. 
It means that our approach finds two qualitatively different ways to violate the specification. Transition $t_{9}$ represents the violation of the obligation 
(sub-formula) $(b \leq 0 \wedge \F_{[0,1]}c \geq 4)$, while $t_{10}$ indicates the violation of the obligation $(b > 0 \wedge \F_{[0,1]} d \geq 6)$ in $\varphi$.
This is in contrast to the falsification testing approach, which stops as soon as the first violation of the specification is detected.

These two experiments reveal another interesting observation -- it is easier to achieve  
full location/transition coverage for $S_{2}$ than the $60\%$-location and $36\%$-transition coverage for $S_{1}$. In fact, the major part of the testing effort goes in 
the attempt to take a transition that is not enabled by any combination of admissible input/output pairs. Finally, we would like to emphasize that adaptive testing is a heuristic --
the procedure can give confidence to the engineer that a transition cannot be enabled, but this indication does not represent 
a formal proof.

\subsection{Case Study}
\label{sec:casestudy}




As case study, we consider the Aircraft Elevator Control System (AECS) proposed in~\cite{mosterman-fdir}. The architecture of AECS has an elevator on the left and right side of the aircraft. Each elevator is equipped with two hydraulic actuators. Either actuator can position the elevator, however at any point in time at most one shall be active. Three different hydraulic systems drive the four actuators. The left (LIO) and right (RIO) outer actuators are controlled by a Primary Flight Control Unit (PFCU1) with a sophisticated input/output control law.  If a failure occurs in the outer actuators or hydraulic systems, a less sophisticated Direct-Link (PFCU2) control law with reduced functionality takes over to handle the left (LDL) and right (RDL) inner actuators. The system uses state machines to coordinate the redundancy and assure its continual fail-operational activity. 

This model has one input variable, Pilot Command, and two output variables, the position of left and right actuators, as measured by the sensors. This is a complex model with
$426$ internal state variables.
When the system behaves correctly, the intended position of the aircraft required by the pilot must be achieved within a predetermined time limit and with a certain accuracy. This can be captured with several requirements. One of them says that whenever the derivative of the Pilot Command $\emph{cmd}$ goes above a threshold $m$ followed by a period of $\tau$ time where the derivative is smaller than $k << m$, the actuator position measured by the sensor must stabilize (become at most $n$ units away from the command signal) within 
$T+t$ time units. This requirement is formalized in IA-STL as the specification $\phi = (\{ \emph{cmd}\}, \{ \emph{lep}\}, \varphi)$, where:
{\small $$
\varphi \equiv \Glob (\emph{cmd'} \leq k  \since_{[\tau,\tau]} \emph{cmd'} \geq m) \rightarrow \F_{[0,T]} \Glob_{[0,t]} (|\emph{cmd} - \emph{pos}| \leq n)) 
$$}
\noindent The symbolic automaton $\aut_{\phi}$ associated to the specification $\phi$ has $241$ locations.
In this case study, we perform four experiments: (1) we compare our adaptive testing approach with falsification testing, (2) we empirically study the effect of 
bounding the total number of simulations on the coverage, (3) we compare the coverage of our adaptive testing approach with random testing and (4) we empirically evaluate the effect of PSO parameters on the coverage.

In the first experiment, we compare our approach to falsification testing in which we use PSO as a global optimizer. We first restrict the adaptive testing procedure to 
have the sink error location as the only target location. For the falsification testing approach, we derived 
from the specification that a violating trace must have at most $40$ samples. Hence, we framed the falsification testing problem as a global optimization problem of finding 
$40$ values that represent an input sequence inducing a run in the specification automaton, which ends in the (sink) error location. For both approaches, 
we set the maximum budget to $12,000$ simulations. The adaptive testing method found a violating behavior by using 
$32$ simulations executed in $121s$ ($69s$ for the initialization of the model and $52s$ for running all simulations). The falsification testing approach could not 
find a violating behavior after $12,000$ simulations executed in $8,854s$. This experiment suggests that the specification can greatly help in incrementally building a violating trace -- the 
structure of the automaton can be used to guide step-by-step the search for the right sequence of inputs.

\begin{figure}[tb]
\centering
\includegraphics[width=0.75\linewidth]{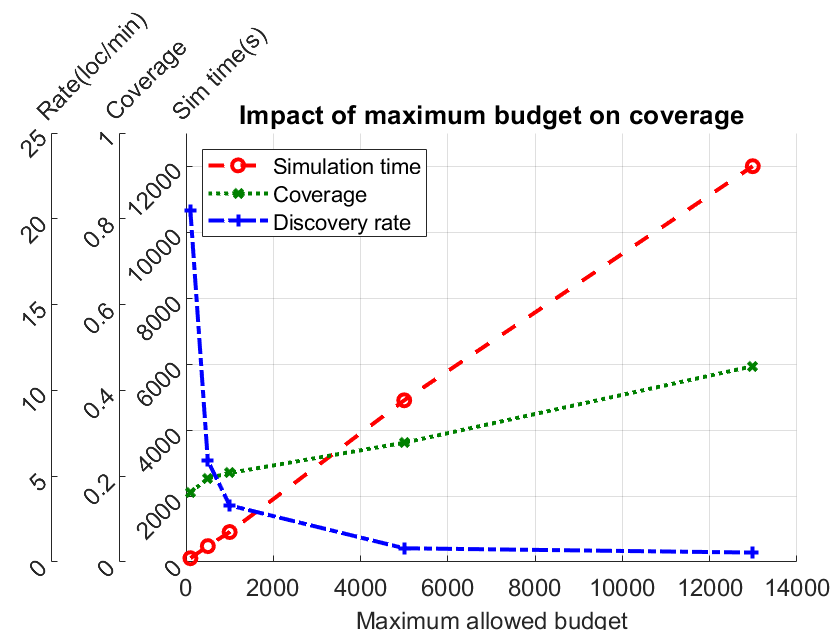}
\caption{Impact of maximum budget on coverage.}
\label{fig:budget}
\end{figure}

In the second experiment, we vary the adaptive testing budget, i.e. the maximum number of simulations that we allow to be used. We evaluate our approach 
with $100$, $500$, $1000$ and $5000$ simulations. In addition, we run the adaptive testing procedure without an upper bound on the budget. The aim 
of this experiment is to study the effect of the budget on coverage.Figure~\ref{fig:budget} summarizes the results of this experiment. 
Without setting the budget, the adaptive testing procedure runs a total of $12986$ 
simulations and visits $110$ out of $241$ locations for the $46\%$ location 
coverage. We can also observe that many of the locations are already discovered 
and visited with a small number of simulations. The figure also shows the 
rate of visiting new locations per minute. We can observe that this rate is rapidly 
dropping with the number of simulations. It is consistent with the testing 
folk theory stating that it is easy to achieve most of the coverage, but that it is difficult 
to discover specific regions. We note that the rate can be used to define 
a stopping criterion for adaptive testing, which represents the desired trade-off 
between the exhaustiveness of testing and the testing effort.


Next, we compare the coverage obtained by adaptive and random testing. 
From the specification, we derived the trace length to be 40 samples. 
Before every simulation, we initialize each sample of the trace with a random value between a predetermined bound, we then run the simulation and record the number of locations covered during the simulation. After completing $12,000$ simulations running for a duration of $10,230s$ random testing covered $10$ out of $241$ locations for $4\%$ location coverage as opposed to  $46\%$ coverage obtained by adaptive testing.

In the fourth experiment, we fixed the total budget to $1,000$ simulations and we varied the two main PSO parameters -- the swarm size and the maximum number of iterations. 
We varied these two parameters by setting their values to $50$, $100$ and $500$, respectively, for a total of $9$ experiments. For each experiment, we measured the 
total execution time of the approach and the number of visited locations. The number of 
visited states varied between $38$ and $50$ for a total execution time that was in the range between $841$ and $960s$. 
It follows that the two PSO parameters have a negligible impact on coverage. Furthermore, the experiments do not indicate  
monotonicity with respect to any of the two parameters. This evaluation suggests a certain stability of the 
adaptive testing approach with respect to the choice of PSO parameters.

%% file: eval_1.pdf_t
\begin{picture}(0,0)%
\includegraphics{eval_1.pdf}%
\end{picture}%
\setlength{\unitlength}{3947sp}%
\begingroup\makeatletter\ifx\SetFigFont\undefined%
\gdef\SetFigFont#1#2#3#4#5{%
  \reset@font\fontsize{#1}{#2pt}%
  \fontfamily{#3}\fontseries{#4}\fontshape{#5}%
  \selectfont}%
\fi\endgroup%
\begin{picture}(5303,3903)(1388,-5380)
\put(3901,-1636){\makebox(0,0)[b]{\smash{{\SetFigFont{12}{14.4}{\rmdefault}{\mddefault}{\updefault}{\color[rgb]{1,0,0}$t_{11}$}%
}}}}
\put(4501,-2311){\makebox(0,0)[b]{\smash{{\SetFigFont{12}{14.4}{\rmdefault}{\mddefault}{\updefault}{\color[rgb]{1,0,0}$t_5$}%
}}}}
\put(4351,-4486){\makebox(0,0)[b]{\smash{{\SetFigFont{12}{14.4}{\rmdefault}{\mddefault}{\updefault}{\color[rgb]{1,0,0}$t_6$}%
}}}}
\put(5026,-3286){\makebox(0,0)[b]{\smash{{\SetFigFont{12}{14.4}{\rmdefault}{\mddefault}{\updefault}{\color[rgb]{1,0,0}$t_7$}%
}}}}
\put(5026,-3586){\makebox(0,0)[b]{\smash{{\SetFigFont{12}{14.4}{\rmdefault}{\mddefault}{\updefault}{\color[rgb]{1,0,0}$t_8$}%
}}}}
\put(6676,-3436){\makebox(0,0)[lb]{\smash{{\SetFigFont{12}{14.4}{\rmdefault}{\mddefault}{\updefault}{\color[rgb]{1,0,0}$t_{13}$}%
}}}}
\put(5776,-3961){\makebox(0,0)[rb]{\smash{{\SetFigFont{12}{14.4}{\rmdefault}{\mddefault}{\updefault}{\color[rgb]{1,0,0}$t_{10}$}%
}}}}
\put(5776,-2911){\makebox(0,0)[rb]{\smash{{\SetFigFont{12}{14.4}{\rmdefault}{\mddefault}{\updefault}{\color[rgb]{1,0,0}$t_9$}%
}}}}
\put(3901,-5311){\makebox(0,0)[b]{\smash{{\SetFigFont{12}{14.4}{\rmdefault}{\mddefault}{\updefault}{\color[rgb]{1,0,0}$t_{12}$}%
}}}}
\put(2851,-3061){\makebox(0,0)[b]{\smash{{\SetFigFont{12}{14.4}{\rmdefault}{\mddefault}{\updefault}{\color[rgb]{0,.56,0}$t_1:467$}%
}}}}
\put(2851,-3811){\makebox(0,0)[b]{\smash{{\SetFigFont{12}{14.4}{\rmdefault}{\mddefault}{\updefault}{\color[rgb]{0,.56,0}$t_2:102$}%
}}}}
\put(3001,-2386){\makebox(0,0)[b]{\smash{{\SetFigFont{12}{14.4}{\rmdefault}{\mddefault}{\updefault}{\color[rgb]{0,.56,0}$t_3:245$}%
}}}}
\put(3001,-4486){\makebox(0,0)[b]{\smash{{\SetFigFont{12}{14.4}{\rmdefault}{\mddefault}{\updefault}{\color[rgb]{0,.56,0}$t_4:39$}%
}}}}
\put(1651,-2611){\makebox(0,0)[b]{\smash{{\SetFigFont{12}{14.4}{\rmdefault}{\mddefault}{\updefault}{\color[rgb]{0,.56,0}$t_0:111$}%
}}}}
\put(5101,-2536){\makebox(0,0)[b]{\smash{{\SetFigFont{12}{14.4}{\rmdefault}{\mddefault}{\updefault}{\color[rgb]{1,0,0}$s_2$}%
}}}}
\put(3601,-3361){\makebox(0,0)[b]{\smash{{\SetFigFont{12}{14.4}{\rmdefault}{\mddefault}{\updefault}{\color[rgb]{0,.56,0}$s_1$}%
}}}}
\put(5101,-4336){\makebox(0,0)[b]{\smash{{\SetFigFont{12}{14.4}{\rmdefault}{\mddefault}{\updefault}{\color[rgb]{0,0,0}$s_3$}%
}}}}
\put(2101,-3361){\makebox(0,0)[b]{\smash{{\SetFigFont{12}{14.4}{\rmdefault}{\mddefault}{\updefault}{\color[rgb]{0,.56,0}$s_0$}%
}}}}
\put(6001,-3436){\makebox(0,0)[b]{\smash{{\SetFigFont{12}{14.4}{\rmdefault}{\mddefault}{\updefault}{\color[rgb]{1,0,0}$s_4$}%
}}}}
\put(2101,-3511){\makebox(0,0)[b]{\smash{{\SetFigFont{12}{14.4}{\rmdefault}{\mddefault}{\updefault}{\color[rgb]{0,.56,0}$887$}%
}}}}
\put(3601,-3511){\makebox(0,0)[b]{\smash{{\SetFigFont{12}{14.4}{\rmdefault}{\mddefault}{\updefault}{\color[rgb]{0,.56,0}$467$}%
}}}}
\end{picture}%

%% file: conc.tex
\section{Conclusions and Future Work}
\label{sec:conc}

We proposed a new adaptive testing approach for covering specifications of CPS. To achieve this goal, 
we combine cooperative games with numerical optimization. Cooperative games use the premise that the tester and the SUT 
are not necessarily adversarial, and that a winning strategy may be possible under the assumption that these two 
entities cooperate. 
We believe that our approach provides novel 
methodological insights on systematic testing of CPS that at the same time aims at effectively falsifying the SUT
in the presence of a fault and providing confidence in the SUT correctness in the absence of a fault. 


%% file: main.bbl
\begin{thebibliography}{10}

\bibitem{AichernigBJKST15}
Bernhard~K. Aichernig, Harald Brandl, Elisabeth J{\"{o}}bstl, Willibald Krenn,
  Rupert Schlick, and Stefan Tiran.
\newblock Killing strategies for model-based mutation testing.
\newblock {\em Softw. Test., Verif. Reliab.}, 25(8):716--748, 2015.

\bibitem{AichernigLN13}
Bernhard~K. Aichernig, Florian Lorber, and Dejan Nickovic.
\newblock Time for mutants - model-based mutation testing with timed automata.
\newblock In {\em Proc. of {TAP} 2013: the 7th International Conference on
  Tests and Proofs}, volume 7942 of {\em LNCS}, pages 20--38. Springer, 2013.

\bibitem{avstl}
Takumi Akazaki and Ichiro Hasuo.
\newblock Time robustness in {MTL} and expressivity in hybrid system
  falsification.
\newblock In {\em Proc. of {CAV} 2015: the 27th International Conference on
  Computer Aided Verification}, volume 9207 of {\em LNCS}, pages 356--374.
  Springer, 2015.

\bibitem{staliro}
Yashwanth Annapureddy, Che Liu, Georgios~E. Fainekos, and Sriram
  Sankaranarayanan.
\newblock {S}-{T}a{L}i{R}o: {A} tool for temporal logic falsification for
  hybrid systems.
\newblock In {\em Proc. of {TACAS} 2011: the 17th International Conference on
  Tools and Algorithms for the Construction and Analysis of Systems}, volume
  6605 of {\em LNCS}, pages 254--257, 2011.

\bibitem{BartocciDDFMNS18}
Ezio Bartocci, Jyotirmoy~V. Deshmukh, Alexandre Donz{\'{e}}, Georgios~E.
  Fainekos, Oded Maler, Dejan Nickovic, and Sriram Sankaranarayanan.
\newblock Specification-based monitoring of cyber-physical systems: {A} survey
  on theory, tools and applications.
\newblock In {\em Lectures on Runtime Verification - Introductory and Advanced
  Topics}, volume 10457 of {\em LNCS}, pages 135--175. Springer, 2018.

\bibitem{BloemFGKPRR19}
Roderick Bloem, G{\"{o}}rschwin Fey, Fabian Greif, Robert K{\"{o}}nighofer,
  Ingo Pill, Heinz Riener, and Franz R{\"{o}}ck.
\newblock Synthesizing adaptive test strategies from temporal logic
  specifications.
\newblock {\em Formal Methods Syst. Des.}, 55(2):103--135, 2019.

\bibitem{BloemHRS15}
Roderick Bloem, Daniel~M. Hein, Franz R{\"{o}}ck, and Richard Schumi.
\newblock Case study: Automatic test case generation for a secure cache
  implementation.
\newblock In {\em Proc. of {TAP} 2015: the 9th International Conference of
  Tests and Proofs}, volume 9154 of {\em LNCS}, pages 58--75. Springer, 2015.

\bibitem{DavidLLN08}
Alexandre David, Kim~Guldstrand Larsen, Shuhao Li, and Brian Nielsen.
\newblock Cooperative testing of timed systems.
\newblock {\em Electron. Notes Theor. Comput. Sci.}, 220(1):79--92, 2008.

\bibitem{Dokhanchi2015}
A.~{Dokhanchi}, A.~{Zutshi}, R.~T. {Sriniva}, S.~{Sankaranarayanan}, and
  G.~{Fainekos}.
\newblock Requirements driven falsification with coverage metrics.
\newblock In {\em 2015 International Conference on Embedded Software (EMSOFT)},
  pages 31--40, 2015.

\bibitem{fainekos-robust}
Georgios~E. Fainekos and George~J. Pappas.
\newblock Robustness of temporal logic specifications for continuous-time
  signals.
\newblock {\em Theor. Comput. Sci.}, 410(42):4262--4291, 2009.

\bibitem{FellnerKSTW17}
Andreas Fellner, Willibald Krenn, Rupert Schlick, Thorsten Tarrach, and Georg
  Weissenbacher.
\newblock Model-based, mutation-driven test case generation via
  heuristic-guided branching search.
\newblock In {\em Proc. of {MEMOCODE} 2017: the 15th {ACM-IEEE} International
  Conference on Formal Methods and Models for System Design}, pages 56--66.
  ACM, 2017.

\bibitem{FerrereNDIK19}
Thomas Ferr{\`{e}}re, Dejan Nickovic, Alexandre Donz{\'{e}}, Hisahiro Ito, and
  James Kapinski.
\newblock Interface-aware signal temporal logic.
\newblock In {\em Proc. of {HSCC} 2019: the 22nd {ACM} International Conference
  on Hybrid Systems}, pages 57--66. {ACM}, 2019.

\bibitem{mosterman-fdir}
Jason Ghidella and Pieter Mosterman.
\newblock Requirements-based testing in aircraft control design.
\newblock In {\em AIAA Modeling and Simulation Technologies Conference and
  Exhibit}, page 5886, 2005.

\bibitem{HenryJM18}
L{\'{e}}o Henry, Thierry J{\'{e}}ron, and Nicolas Markey.
\newblock Control strategies for off-line testing of timed systems.
\newblock In {\em Proc. of {SPIN} 2018: the 25th International Symposium on
  Model Checking Software}, volume 10869 of {\em LNCS}, pages 171--189.
  Springer, 2018.

\bibitem{JiaH11}
Yue Jia and Mark Harman.
\newblock An analysis and survey of the development of mutation testing.
\newblock {\em {IEEE} Trans. Software Eng.}, 37(5):649--678, 2011.

\bibitem{pso1995}
J.~{Kennedy} and R.~{Eberhart}.
\newblock Particle swarm optimization.
\newblock In {\em Proceedings of ICNN'95 - International Conference on Neural
  Networks}, volume~4, pages 1942--1948 vol.4, 1995.

\bibitem{MalerN13}
Oded Maler and Dejan Nickovic.
\newblock Monitoring properties of analog and mixed-signal circuits.
\newblock {\em Int. J. Softw. Tools Technol. Transf.}, 15(3):247--268, 2013.

\bibitem{TanSL04}
Li~Tan, Oleg Sokolsky, and Insup Lee.
\newblock Specification-based testing with linear temporal logic.
\newblock In {\em Proc. of {IRI} 2004: the 2004 {IEEE} International Conference
  on Information Reuse and Integration}, pages 493--498. {IEEE} Systems, Man,
  and Cybernetics Society, 2004.

\bibitem{Tretmans08}
Jan Tretmans.
\newblock Model based testing with labelled transition systems.
\newblock In {\em Formal Methods and Testing, An Outcome of the {FORTEST}
  Network, Revised Selected Papers}, volume 4949 of {\em LNCS}, pages 1--38.
  Springer, 2008.

\bibitem{Yannakakis2004}
M.~{Yannakakis}.
\newblock Testing, optimization, and games.
\newblock In {\em Proc. of LICS 2004: the 19th Annual IEEE Symposium on Logic
  in Computer Science, 2004.}, pages 78--88. IEEE, 2004.

\end{thebibliography}
